# *Saccharina latissima*, candy-factory waste, and digestate from full-scale biogas plant as alternative carbohydrate and nutrient sources for lactic acid production

Eleftheria Papadopoulou[a]; Charlene Vance[b]; Paloma S. Rozene Vallespin[a]; Panagiotis Tsapekos[a]; Irini Angelidaki[a*]

[a] Department of Chemical and Biochemical Engineering, Technical University of Denmark, Kgs. Lyngby DK-2800, Denmark

[b] School of Biosystems & Food Engineering, University College Dublin, Agriculture Building, UCD Belfield, Dublin 4, Ireland

*Corresponding author:

E-mail address: iria@kt.dtu.dk (I. Angelidaki)

Postal address: Department of Chemical and Biochemical Engineering, Technical University of Denmark, Kgs. Lyngby DK-2800, Denmark

**Abstract**

To substitute petroleum-based materials with bio-based alternatives, microbial fermentation combined with inexpensive biomass is suggested. In this study *Saccharina latissima* hydrolysate, candy-factory waste, and digestate from full-scale biogas plant were explored as substrates for lactic acid production. The lactic acid bacteria *Enterococcus faecium*, *Lactobacillus plantarum*, and *Pediococcus pentosaceus* were tested as starter cultures. Sugars released from seaweed hydrolysate and candy-waste were successfully utilized by the studied bacterial strains. Additionally, seaweed hydrolysate and digestate served as nutrient supplements supporting microbial fermentation. According to the highest achieved relative lactic acid production, a scaled-up co-fermentation of candy-waste and digestate was performed. Lactic acid reached a concentration of 65.65 g/L, with 61.69% relative lactic acid production, and 1.37 g/L/hour productivity. The findings indicate that lactic acid can be successfully produced from low-cost industrial residues.

# 1  Introduction

Plastic was introduced to the market around 1950 (Geyer et al., 2017), and it is now an integral part of everyday life. A study conducted by Lebreton and Andrady (Lebreton and Andrady, 2019), appreciated that 155-265 Mt/year of mismanaged plastic waste is expected to be produced globally, by 2060. Thus, European Union announced the "European strategy for plastics in a circular economy" action plan (European Commission, 2018), in January 2018. This strategy aims to reduce, recycle, and transform the way that plastic is produced.

Considering the transformation of plastic production technologies, the idea of bio-degradable plastics has arisen and has claimed the position of petroleum-based derivatives (Alexandri et al., 2022). Bioplastics are produced by the polymerization of poly-lactic acid (PLA), synthesized initially by lactic acid (LA), an organic acid used as a building block. A significant amount (90%) of lactic acid utilized for industrial use, is nowadays a derivative of microbial fermentations, rather than chemical synthesis (Alves de Oliveira et al., 2018). The most promising microbial group participating for LA fermentation processes are lactic acid bacteria (LAB) (Bosma et al., 2017).

In search of robust microorganisms with low nutritional needs and able to metabolize a variety of carbohydrates, the isolation of autochthonous LAB with niche-specific characteristics and new properties is required (Ruiz Rodríguez et al., 2019). *Enterococcus* species are LAB with wide distribution in nature and high biotechnological interest, studied for the production of secondary metabolites, such as lactic acid (Selim et al., 2021; Unban et al., 2020). Specifically, *Enterococcus faecium* strains have been used for the fermentation of residue substrates. For example, fermentation of corn steep water, resulted to 44.60 g/L LA from 60.00 g/L initial sugars (Selim et al., 2021). Additionally, 93.07 g/L LA were generated from 125.70 g/L initial gelatinized starchy waste from rice noodle factory (Unban et al., 2020).

For an optimized process, the selected microorganism should be combined with an abundant and inexpensive feedstock. Lignocellulosic biomass has been previously used for LA fermentation (Cubas-Cano et al., 2020). However, the exploitation of lignocellulosic material requires the application of harsh pre-treatment strategies, increasing the final fermentation cost (Komesu et al., 2017). Thus, substrates with non- or low lignin content, such as seaweed (Marinho et al., 2016) and organic residues (Thygesen et al., 2021) are proposed as alternative feedstocks for the production of value-added products.

Seaweed is regarded among others as a sustainable biorefinery feedstock (Nakhate and van der Meer, 2021), due to the low nutrient requirements for aquaculture cultivation, and the numerous biotechnological properties. Characteristics, such as high carbohydrate content (up to 60% of dry biomass), lack of lignin, and abundance in bio-compounds, set seaweed as a potential substrate (Marinho et al., 2016). Due to the aforementioned reasons, macroalgae have been used as feedstock for the generation of biorefinery products (i.e., acetone, butanol, and ethanol) (Schultze-Jena et al., 2022).

Municipal (Thygesen et al., 2021), or industrial (Alexandri et al., 2022) organic residues have also magnetized scientific interest as potential fermentation substrates for several reasons. Firstly, as 230 Mt/year of municipal and industrial waste are estimated to be generated in European Union (Alibardi et al., 2020), management planning is considered crucial. Secondly, depending on the biomass source, these substrates can be rich in carbohydrates (Vigato et al., 2022), proteins (Zhao et al., 2022), or other nutrients such as nitrogen and phosphorus (Wang et al., 2022), valuable for microbial growth. Finally, lignin-poor residues, such as source-sorted organic household waste do not require harsh pre-treatment before the hydrolysis step (Thygesen et al., 2021). The combination of different residues for the generation of a substrate with desirable characteristics, has been mainly applied in anaerobic digestion (AD) processes (Negro et al., 2022). However, it could be also considered as a potential strategy for lactic acid production.

The scope of this project was to study and compare three sustainable carbohydrate and/or nutrient sources, such as seaweed, candy-factory waste, and digestate as alternative options for glucose as carbon source and de Man Rogosa Sharpe (MRS) medium as nutrient supply, for LAB growth. The three biomasses were characterized for sugar, and nitrogen content. Pre-treatment strategies were applied for the optimization of the fermentation process. LAB *E. faecium* was chosen initially as a starter culture, but strains of the species *Pediococcus pentosaceus* and *Lactobacillus plantarum* were also tested as potential fermentation candidates. Based on the pretreatment strategies and yields achieved by the flask fermentation trials, a preliminary economic analysis was conducted to estimate the cost in each of the proposed scenarios. According to the collected data, an optimized fermentation substrate, and a suitable bacterial strain were combined inside 5-L bioreactor batch fermentations.

## 2 Materials and Methods

### 2.1 Microbial strains and chemical reagents

Three seaweed isolates (Papadopoulou et al., 2023) were used as LAB strains. *Enterococcus faecium* (GenBank accession no. **SAMN32411886**) was used as starter culture for the selection of the optimum sugar and nutrient source. LAB *Pediococcus pentosaceus* (GenBank accession no. **SAMN32411881**), and *Lactobacillus plantarum* (GenBank accession no. **SAMN32411876**) were also tested as potential microbial factories using real-life substrates. All bacterial candidates were DL-lactic acid producers. *P. pentosaceus* and *L. plantarum* produced 1.06-, and 1.32-fold higher amount of L- compared to D-lactic acid, respectively. On the contrary *E. faecium* produced 1.59-fold more D-, rather than L-lactic acid. The bacterial isolates were stored in 20% glycerol stocks, at -80˚C. Fermentation inocula were prepared in the common synthetic media for LAB growth (De MAN et al., 1960). Glucose was added externally with a Minisart®, 0.22 μm pore size syringe filters (Fisher Scientific®, Roskilde, Denmark), to avoid potential degradation during the autoclaving process. The seed cultures were prepared in anaerobically sealed serum bottles (250 ml) with working volume 50 ml, at pH 6.5, and 37˚C incubation temperature, for 12 hours, without shaking. All chemicals and enzymes used in this study were of analytical grade, purchased from MERCK A/S (Søborg, Denmark).

### 2.2 Feedstock collection and characterization

Two biomasses were suggested as potential sugar sources for lactic acid fermentation, brown seaweed *Saccharina latissima* and candy-factory waste (denoted as candy-waste). *S. latissima* and digestate from full-scale biogas plant were proposed as alternative nutrient sources. *S. latissima* biomass was collected from an aquaculture handled by Hjarnø Havbrug A/S (Horsens, Denmark) between 2013-2014. Biomass was received by a previous study (Marinho et al., 2016). Candy-waste was obtained by Trolli Ibérica A/S (Paterna, Spain) and stored at -20˚C. Digestate was collected from Hashøj biogas plant (Dalmose, Denmark), and consisted of 80% animal manure and 20% food or industrial waste. Digestate was stored anaerobically, at 54 ± 1˚C.

**2.3    Feedstock preparation and flask fermentation**

Mixed *S. latissima* biomass collected in different months was hydrolyzed (Papadopoulou et al., 2023) before fermentation. Hydrolysate was tested as sugar- and nutrient source for lactic acid production. When seaweed was tested as sugar source, MRS components were added in the substrate, whereas when seaweed was tested as nutrient source, external chemicals were not included. The solution was autoclaved at 121˚C for 15 minutes. Due to the low sugar content of fresh seaweed (12.21 ± 0.79% DM) and consequently of seaweed hydrolysate (11.80 ± 0.08 g/L), glucose was added as an external sugar source to obtain a final initial sugar concentration of 30.00 g/L in the fermentation media. When glucose was added, the produced lactic acid was excluded from the final calculations.

Digestate was studied only as nutrient source. It was manually treated with a 2 mm pore size sieve to remove large particles. Treated digestate was studied undiluted, or diluted 1:1, 1:3, 1:5, and 1:9 (digestate: $H_2O$) to detect the optimal concentration for LAB growth. Furthermore, three sterilization methods were applied: autoclaving at 121˚C for 15 minutes, pasteurization at 70˚C for 30 minutes, or filtration with a combination of Minisart®, 0.45 μm and 0.22 μm pore size sterile filters (Fisher Scientific®,Roskilde, Denmark). Nutrient availability of digestate was tested in presence of MRS components, without, or in presence of 8.00 g/L beef extract.

Finally, candy-waste was homogenized and tested as carbohydrate source, with the addition of MRS components, and as nutrient source without chemical addition. Once the most suitable LAB was detected, the dilution ratio between candy-waste: digestate was investigated.

Batch fermentations with the different substrates were conducted in anaerobically sealed serum bottles (250 ml) of 50 ml working volume, pH 6.55 ± 0.27, inoculated with 2.5% (v/v) *E. faecium* and incubated at 37˚C. The initial PTS sugar concentration (glucose, sucrose, mannitol) (McCoy et al., 2015) was set at 31.00 ± 4.82 g/L for all experiments. The fermentation was completed when pH reached values between 3.2-4.0. Manual shaking was applied during the sampling points. Batch fermentations prepared with glucose and MRS nutrients as fermentation media were used as control for the study. The samples were analyzed for organic acids and sugar content.

**2.4 Economic analysis**

A simplified economic assessment was performed to evaluate the production cost of lactic acid fermentation, considering different substrates. Three options were considered: 1) glucose as carbon source and MRS media as nutrient source, 2) glucose as additional carbon source and seaweed hydrolysate as nutrient source, and 3) candy waste as carbon source and digestate as nutrient source. For comparability, the data was adjusted on a uniform starting sugar concentration of 30 g/L. The inputs from the flask experiments were then scaled up to reflect the volumes needed for the 5-L validation experiment. The costs for feedstocks and their transportation, chemicals for pre-treatments methods, labor, water, and energy demand were included. The production cost of lactic acid fermentation was thus calculated according to Equation a.

*Equation a. Production cost (DKK) = total feedstock cost + total chemicals cost + total labor cost + total water cost + total energy cost*

To perform the assessment, chemical costs (glucose, MRS components, citric acid, enzymes) were collected from MERCK A/S (Søborg, Denmark). The candy waste and digestate did not have a purchase cost. However, transport was considered with a cost of 2.74 DKK/t/km (van der Meulen et al., 2020). Transport distance was assumed to be 26 km and 20 km for candy waste and digestate, respectively, based on real distances from a candy factory (Hvidovre, Denmark) and a biogas plant (Dalmose, Denmark) located in Denmark, to Technical University of Denmark (DTU, Kgs. Lyngby). Seaweed cost was estimated to be 0.001 DKK/g fresh weight (van den Burg et al., 2019) excluding transport cost. The transportation was assumed 250 km from a real seaweed farm (Horsens, Denmark) to Technical University of Denmark (DTU, Kgs. Lyngby). The cost of water was assumed to be 74 DKK/m$^3$ (danva.dk, 2023), and the cost of electricity was assumed to be 2.4 DKK/kWh (www.globalpetrolprices.com, 2023). Finally, the cost of labor was set 139.42 DKK/hour (www.salaryexplorer.com, 2023), considering the average salary provided to a laboratory technician in Denmark.

## 2.5 Up-scaled validation experiments

Validation experiments were carried out in pH-constant 5-L bioreactors (BioBench, Biostream International BV, Doetinchem, The Netherlands). Three replicated bioreactors were operated with the chosen real-life substrate (MERCK A/S, Søborg, Denmark), and one bioreactor run as control with synthetic MRS media. The bioreactors were autoclaved, including the real-life nutrient source, at 121°C for 15 minutes. The carbohydrate source and 5% (v/v) inoculum were added under sterile conditions. The system was flushed with $N_2$ to create anaerobic conditions. The bioreactors were operated at 37°C, stirring at 100 rpm, pH was controlled automatically to 6.55 ± 0.27 with NaOH 8.00 M for 48 hours. Samples were collected every 2 hours for the first 12 hours, and every 12 hours until the end of the process for the analysis of sugars and organic acids.

## 2.6 Analytical methods

Total solids (TS), volatile solids (VS), total Kjeldahl nitrogen (TKN), and total ammonia nitrogen ($NH_4^+$-N) were measured according to standard methods for the examination of water and wastewater (Baird et al., 2017). Organic acids and glucose were analyzed with High Performance Liquid Chromatography (HPLC), equipped with a refractive detector, and a Bio-Rad HPX-87H (300 mm × 7.8 mm) column. Eluent was 12 mM $H_2SO_4$, with a flow rate of 0.60 ml/minute. The column oven temperature was set at 63°C. Sugar content was measured by Ion Chromatography (IC), Dionex ICS-6000 HPIC System. The System was equipped with a Dionex™ CarboPac™ PA20 column (3 × 150 mm). The eluent was a mix of 10-100 mM NaOH, set at a flow rate of 0.4 ml/minute. The column oven temperature was set at 30°C for 35 minutes. Samples analyzed with both HPLC and IC were firstly centrifuged at 10000 rpm for 10 minutes, diluted according to the machine range, and filtered through non-sterile 0.22 μm pore size filters (MERCK A/S, Søborg, Denmark). Elemental analysis (EA) for the quantification (%) of nitrogen (N), and carbon (C) was conducted with EuroEA3000 CHNS-O Analyzer (EVISA, EuroVector S.p.A., Milan, Italy). Metal content of digestate was determined with Inductively Coupled Plasma (ICP) spectroscopy (Optical Emission Spectrometer, Model Optima 4300 Dv, PerkinElmer, USA).

## 2.7 Statistical analysis

Statistical analysis and graphic representation were prepared in GraphPad Prism 9.3.1. All values represent means of three repetitions, except if differently stated. Error bars depict standard deviation (SD). One-way analysis of variance (One-way ANOVA) and a multiple comparison test (Tukey's Test) were applied to test significance. An asterisk represents significantly different results (P ≤0.05). Relative lactic acid production (%) was appreciated according to the equation below. Practical LA concentration (g/L) depicts the concentration detected by the experimental analysis. The theoretical LA concentration was calculated according to the chemical equation and molar mass of each reactant. The theoretical yield of lactic acid on glucose is 1 $g_{-LA}/g_{-Glu}$, on mannitol 0.99 $g_{-LA}/g_{-Man}$, on sucrose 1.05 $g_{-LA}/g_{-Suc}$, and on maltose 1.05 $g_{-LA}/g_{-Mal}$.

$$Equation\ b \quad Relative\ LA\ production = \frac{Practical\ LA\ concentration}{Theoretical\ LA\ concentration} \times 100$$

## 3 Results and Discussion

### 3.1 Candy-waste, and seaweed hydrolysate as alternative carbon sources

Two feedstocks were proposed as alternative carbon sources for LA fermentation, *S. latissima* hydrolysate and candy-waste. Sugar analysis revealed that the total sugar content of candy-waste was 23.66-fold higher than seaweed hydrolysate. Furthermore, both feedstocks consisted of different combination of fermentable sugars. Glucose (8.31 ± 0.15 g/L), and mannitol (2.97 ± 0.61 g/L) were the most abundant fermentable sugars in seaweed hydrolysate, whereas glucose (50.25 ± 3.57 g/L), sucrose (100.71 ± 3.13), and maltose (128.23 ± 1.74 g/L) were detected in candy-waste.

Batch fermentations were conducted using either pure glucose, or each of the two feedstocks as carbon source (Figure 1, a). MRS components were added as nutrient source. Fermentations happened under identical conditions, using *E. faecium* inoculum. Tests on pure glucose revealed the highest relative production (90.19 ± 0.20%). Comparing the relative production, seaweed hydrolysate was associated with the second most promising performance (85.21 ± 0.27%), followed by candy-waste (24.61 ± 0.44%).

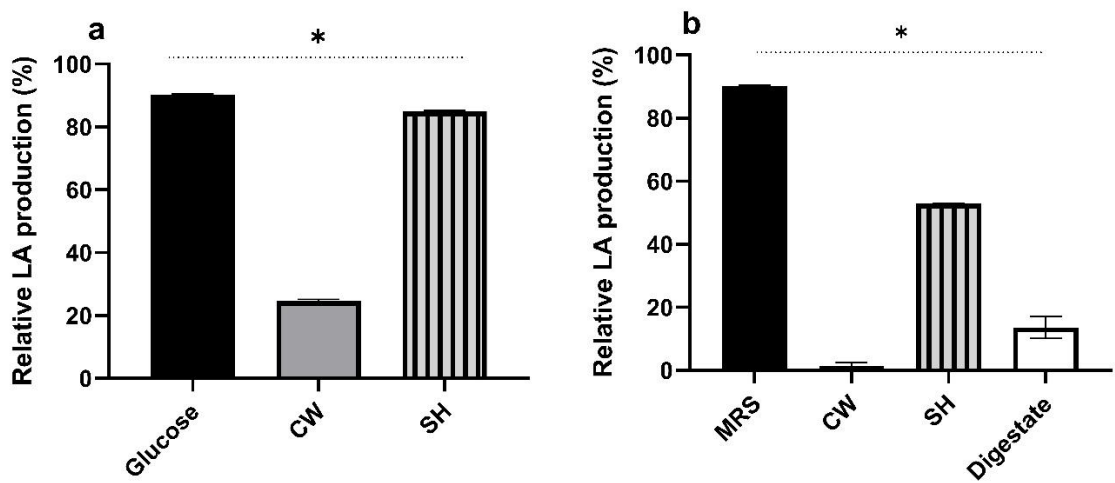

Figure 1. Sugar (a) and nutrient (b) sources selection for lactic acid fermentation. Glucose, candy-waste (CW), and seaweed hydrolysate (SH) were used as carbon sources. Synthetic media (MRS), candy-waste (CW), seaweed hydrolysate (SH), and digestate were used as nutrient sources. Values are mean (n=3) ± standard deviation. The asterisk symbol shows statistical significance (P ≤0.05).

The high relative production in *S. latissima* hydrolysate proved the value of this biomass as biorefinery feedstock. However, the low sugar content of the specific biomass was disincentive for use in industrial processes. Sugar content of *S. latissima* has been found to vary seasonally (Marinho et al., 2016). Marinho *et al.* found that the highest amount of glucose (10.15-24.88% DM) and mannitol (maximum 10.3% DM) was noticed in summer months (May-September). The seaweed included in this study was a mix of different periods and had noticeable lower amount of fermentable sugars. However, preservation of brown kelp with high sugar content could overcome the barriers of space and time (Sandbakken et al., 2018).

Candy-waste showed the lowest performance. However, the initial concentration was not optimized. Sugar overload may have caused substrate inhibition on the microorganism (Cubas-Cano et al., 2018). On the other hand, the high amount of fermentable sugars in this feedstock triggered further research. Furthermore, even though candy-waste may have different characteristics depending on raw materials and provider company, it is not strongly affected by seasonality, as does seaweed.

Seaweed isolates *E. faecium*, *P. pentosaceus*, and *L. plantarum* were tested on both feedstocks as seed (Figure 2). *P. pentosaceus* achieved the highest (88.55 ± 2.28%) and *E. faecium* the lowest (48.85 ± 2.92%) relative production in seaweed hydrolysate. However, *L. plantarum* reached by 1.35- and 1.63-fold higher relative production, compared to *E. faecium* and *P. pentosaceus*, respectively when cultivated with candy-waste. All bacterial strains had significantly better performance using seaweed hydrolysate, rather than candy-waste. This result occurred probably due to adaptation of these strains to seaweed biomass (Ruiz Rodríguez et al., 2019), since they are autochthonous flora of brown kelp.

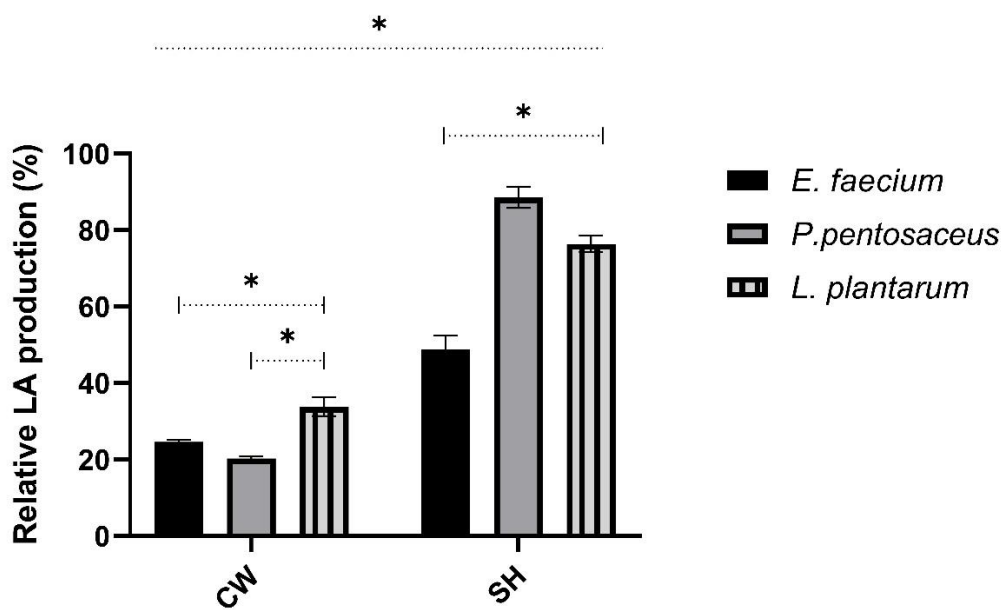

Figure 2. Enterococcus faecium, Pediococcus pentosaceus, and Lactobacillus plantarum were tested as starter cultures for fermentation of candy-waste (CW) and seaweed hydrolysate (SH), as sugar sources. Values are mean (n=3) ± standard deviation. The asterisk symbol shows statistical significance (P ≤0.05).

A study published by Nagarajan *et al.* (Nagarajan et al., 2022) tested *Lactobacillus sp.* and *Weissella sp.* strains for lactic acid production in 40.00 g/L glucose. The results showed that *L. plantarum* reached the highest LA titer (22.35 g/L) and yield (0.75 $g_{-LA}/g_{-glucose}$) compared to the rest of the tested strains. LA yields were fluctuating between 0.59-0.75 $g_{-LA}/g_{-glucose}$, depending on the bacterial species and specific strain. In the same study, seaweed hydrolysates from green, red and brown macroalgae were also tested. The variations in the yields of the same strain (*L. plantarum* 23) ranged from 0.75-0.89 $g_{-LA}/g_{-sugars}$ on different feedstocks and initial sugar concentration, highlights the fact that a specific microorganism changes metabolism depending on the carbon source and the adaptation to the specific feedstock.

### 3.2 Alternative nutrient sources for LA production

Seaweed hydrolysate and digestate were tested as substitutes for MRS components (Figure 1, b). In both treatments pure glucose was the sugar source. Sugar content of the seaweed hydrolysate was taken under consideration. Candy-waste was tested also as nutrient input (Figure 1, b). Synthetic media promoted the highest relative production (90.19 ± 0.20%), in comparison with seaweed hydrolysate (52.83 ± 0.18%), and undiluted digestate (13.58 ± 2.80 %). Candy-waste did not act as nutrient source (1.43 ± 0.84%), as it is a product of gelatin, mixed sugars, flavor and preserving factors, lacking nutrient sources for bacterial growth (Zavistanaviciute et al., 2022).

Relative lactic acid production showed that even though both seaweed hydrolysate and digestate could serve as nutrient sources, seaweed prevailed upon digestate. Seaweed has been characterized as a substrate with high variety of bio-compounds, such as minerals, vitamins, and proteins (Yang et al., 2021). On the other hand, liquid-digestate residues are rich in minerals and nutrients such as nitrogen in the form of ammonia-N and phosphorus as phosphate (Duan et al., 2020; Wang et al., 2022).

### 3.3 Digestate optimization and candy-waste: digestate ratio for LA fermentation

To optimize digestate performance, avoiding degradation of bio-compounds, three sterilization strategies were tested: autoclaving, pasteurization, and serial filtration. The filtration was unsuccessful due to feedstock consistency. No differences, considering relative production, were observed between sterilization and pasteurization methods (Figure 3, a).

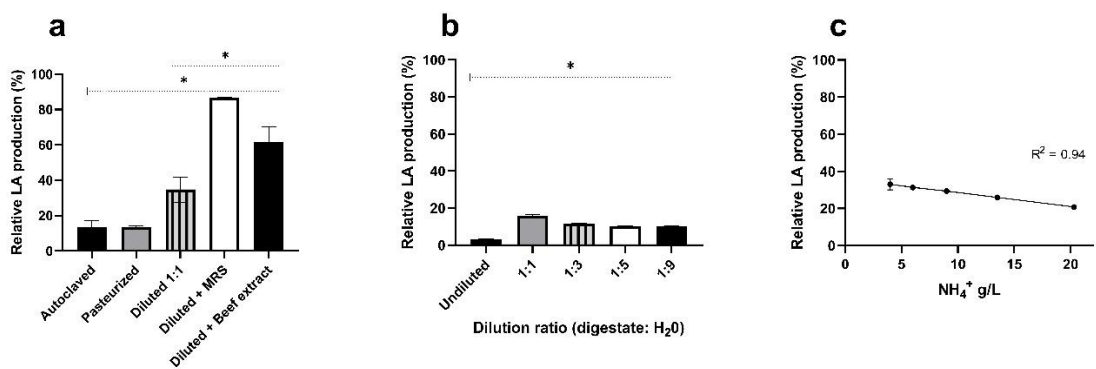

Figure 3. Digestate optimization: sterilization strategy and nutrient optimization (a), optimum dilution of digestate (b), effect of $NH_4^+$ on LA production (c). Values are mean (n=3) ± standard deviation. Pasteurized condition is mean (n=2) ± standard deviation, and ANOVA was not applied. The asterisk symbol shows statistical significance (P ≤0.05).

Digestate was also diluted 1:1 (digestate: $H_2O$) to detect the existence of potential inhibiting compounds. The dilution improved the relative LA concentration by 2.56-fold. Heavy metals, and other organic micro-pollutants (Sfetsas et al., 2022), such as heat resistant antibiotics (Durand et al., 2019) could be inhibitors of bacterial growth. A previous study (Ameen et al., 2020) on LAB isolated by seaweed, showed increased resistance to heavy metals and antibiotics spread in the environment. However, bacterial growth was still affected and especially the existence of $Cd^{2+}$ and $Pb^{2+}$ was lethal for the microorganisms.

Lack of nutrients was tested adding all MRS components, or only beef extract in the diluted digestate (Figure 3, a). Diluted digestate with MRS depicted the highest improvement (86.70 ± 0.10%). However, beef extract also upgraded the relative production by 1.77-fold. The improvement of digestate with beef extract indicated potential lack of proteins in the feedstock. This hypothesis was supported by the low amount of organic carbon (0.24 ± 0.08 g/L) in the digestate, and by the fact that only the liquid part of digestate was used. The liquid fraction of digestate is mainly rich in nitrogen and phosphorus (Sfetsas et al., 2022). Except for proteins, beef extract is rich in nucleotides, vitamins, and trace elements which may have been consumed by acidogenic microorganisms during AD (Meegoda et al., 2018). Thus, to substitute MRS nutrients digestate dilution and addition of beef extract was proposed.

The optimal digestate:$H_2O$ dilution ratio was further investigated, using candy-waste as carbon source (Figure 3, b). The findings proved that the highest relative production (15.99 ± 0.43%) was depicted at 1:1 dilution. This value decreased with further dilutions. The results were not statistically significant ($P \leq 0.05$). The relative production did not fit previously presented results (Figure 1), as the optimal candy-waste concentration was not determined. To explain the decrease of relative production, the $NH_4^+$ concentration enhancing LA production was also tested (Figure 3). The results showed that concentrations between 4.00-6.00 g/L $NH_4^+$ led to optimum LA concentration. The initial total nitrogen concentration in digestate was estimated to be 3.28 ± 0.05 g/L. Consequently, further dilutions led to lower $NH_4^+$ concentration in the fermentation media, inhibiting bacterial production.

Finally, the optimum candy-waste:digestate ratio for *E. faecium,* was detected with serial dilutions (Figure 4). Digestate diluted 1:1 and 8.00 g/L meat extract, was used as nutrient source. Undiluted candy-waste did not support growth, due to high solute viscosity. The highest relative production was 87.31 ± 1.70%, at the lowest sugar concentration (30.15 g/L). Relative production decreased to 18.42 ± 0.27% as the sugar concentration increased (114.55 g/L). *L. plantarum* growing in the same conditions, showed similar fluctuations (17.31 ± 1.20%, at 114.55 g/L initial sugars).

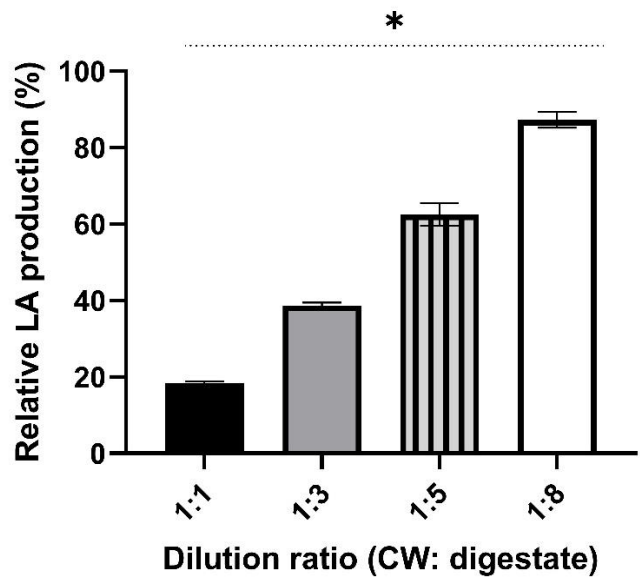

Figure 4. Dilution ratio of candy-waste: digestate. Values are mean (n=3) ± standard deviation. The asterisk symbol shows statistical significance (P ≤0.05).

### 3.4 Economic analysis of lactic acid fermentation

A simplified economic analysis was applied considering three conditions: 1) glucose and MRS nutrients, 2) glucose and seaweed hydrolysate, and 3) candy-waste and digestate, based on a 5-L fermenter volume (Figure 5). Total cost was higher in condition 1 (2491.27 DKK) and condition 2 (3083.40 DKK), compared to condition 3 (1144.82 DKK). Utilization of MRS nutrients increased the input costs (97% of total cost). On the other hand, the biomass pre-treatment and substrate preparation increased the labor cost by 41% and 46% of total cost, respectively.

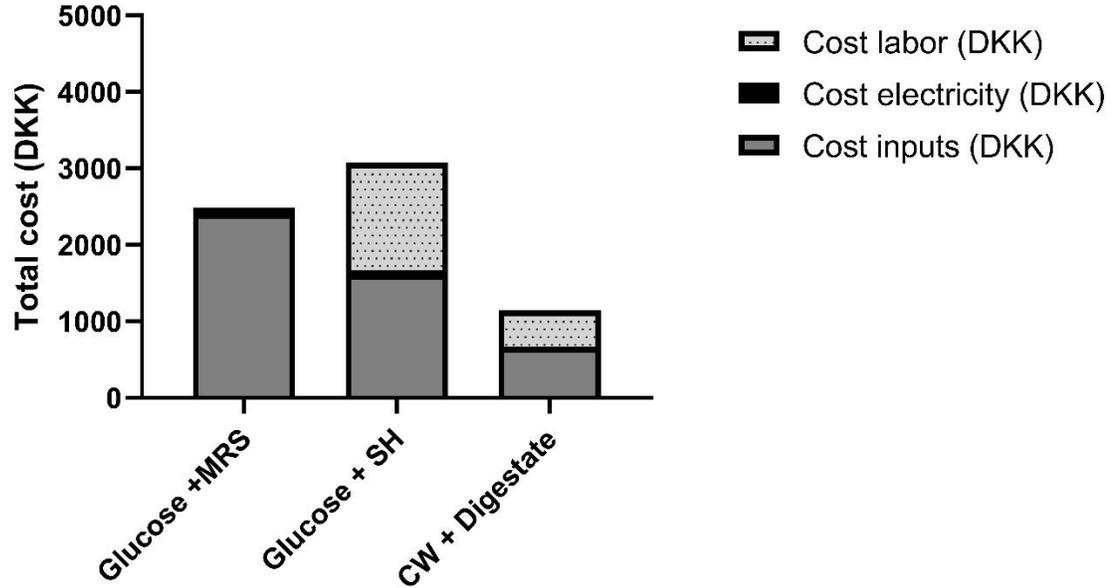

Figure 5. Total production cost for lactic acid fermentation using three different substrates: a. glucose + MRS, b. glucose + seaweed hydrolysate (SH), c. candy-waste (CW) + digestate. The cost of labor, electricity, and inputs was considered in each case.

This preliminary economic assessment focused on the comparison of feedstocks and pre-treatment methods for lab-scale fermentation, excluding the costs of the down-streaming process. Thus, production costs are not comparable to commercial scale techno-economic assessments. However, the cost distribution reflects similar trends to commercial scale. Kwan *et al.* (Kwan et al., 2015) found that 75% of operational costs of lactic acid production were attributed to the production of raw materials, of which glucose (carbon source) made up 68% and the cultivation medium (nutrient source) made up 32%. In another study (Munagala et al., 2021), where hydrolyzed sugarcane bagasse was presumed as the feedstock, it was found that the pre-treatment and hydrolysis steps contributed to 83% of total costs. This study also determined a lactic acid production price of 2.9 USD/kg, which is two- to three-fold higher than production prices from conventional methods. Finally, labor costs were found to contribute to 12-18% of total annual costs in commercial-scale operations (Kwan et al., 2018, 2015), where 70% of all labor demand was concentrated in fermentation processes (Kwan et al., 2018).

## 3.5 Experimental validation in 5-L bioreactors

According to the findings, three replicated bioreactors with real-life substrate were prepared with candy-waste and digestate. The dilution ratio was 1:1 (candy-waste: digestate), to test the improvement of scaling-up. One bioreactor was prepared with glucose and MRS nutrients as positive control. All bioreactors operated under the same conditions. The fermentation was terminated after 48 hours, when all the sugars in the control bioreactor were consumed.

Glucose was firstly consumed after 8 hours, and sucrose after 12-24 hours (Figure 6). At the end of the fermentation 39.25 ± 3.83 g/L maltose remained in the broth. In the presence of multiple carbon sources glucose is preferred by most organisms and often blocks the metabolization of other secondary sugars through the inducer exclusion (Jeckelmann and Erni, 2020). This phenomenon has been studied extensively among others in enterobacteria, where the introduction of glucose in the bacterial cells activates the formation of the dephosphorylated EIIA$^{Glc}$ enzyme, which inhibits the transportation of maltose in the cell (Monedero et al., 2008).

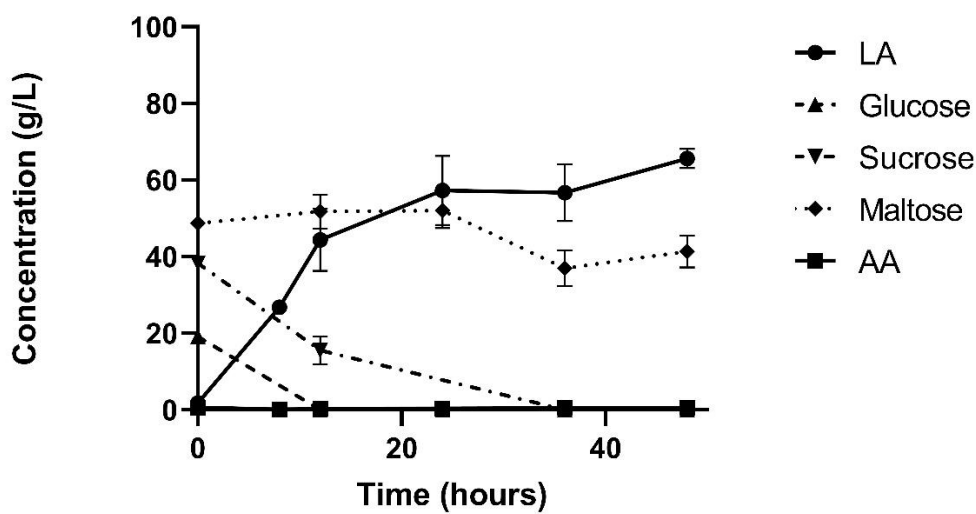

Figure 6. Organic acids (lactic and acetic) production and sugar consumption during co-fermentation of candy-waste and digestate with Lactobacillus plantarum. Values are mean (n=3) ± standard deviation. Sucrose and maltose values are mean (n=2) ± standard deviation.

Metabolic products such as ethanol, propionate, iso-butyrate, butyrate and iso-valerate were detected in amounts less than 0.10 g/L, while valerate and hexanoate were not detected. Acetate was detected in concentrations less than 1.00 g/L, with relative production increasing to 0.51 ± 0.06% after 48 hours of fermentation. LA reached a maximum of 65.65 ± 2.05 g/L, an amount almost identical with the 65.80 g/L LA produced in the control bioreactor. The relative LA production was 61.69 ± 3.41%, and the productivity 1.37 ± 0.04 g/L/hour, compared to 97.86% and 1.37 g/L/hour achieved in the control bioreactor, at the same time point. The relative LA production increased 3.58-fold, compared to the flask fermentations with the same initial sugar concentration.

The LA titer and relative production found in this study support the idea of co-fermentation between digestate and candy-waste. The selected digestate was the eluent of digestion of a mixture of 80% animal manure and 20% food or industrial waste. Most of degradable organic matter in AD is mineralized to biogas. Therefore, only minor organic degradable matter remains in the digestate, which is mainly in the form of volatile fatty acids. Waste streams have been previously applied as fermentation substrates for lactic acid production (Alexandri et al., 2022; Alvarado-Morales et al., 2021; Neu et al., 2016; Pleissner et al., 2017; Zhang et al., 2022). Zhang *et al.* (Zhang et al., 2022) obtained 39.30 g/L lactic acid with a corresponding yield of 0.75 $g_{-LA}/g_{-sugars}$ from 61.70 g/L initial sugar concentration. Pleissner *et al.* (Pleissner et al., 2017) achieved final lactic acid titer of 60.50 g/L with yield 0.64 $g_{-LA}/g_{-sugars}$. In later studies (Pleissner et al., 2021) digested sewage sludge was proposed as potential source for lactic acid fermentation, provided that it is combined with carbon-rich feedstocks. In another study where coffee mucilage was used as fermentation substrate and *Bacillus coagulans* as starter culture more than 40 g/L lactic acid were produced by 60 g/L free sugars (Neu et al., 2016). Lignocellulosic hydrolysates were also applied as substrate for the co-cultivation of *Lactobacillus coryniformis* and *Leuconostoc spp.* to produce 21.7 g/L lactic acid (Alexandri et al., 2022). Furthermore, municipal biopulp was characterized as an interesting feedstock, generating 16.1 g/L lactic acid with a yield of 82.6% g/g of total sugars (Alvarado-Morales et al., 2021). Co-fermentation of biomasses with different characteristics has proved to enhance lactic acid production and simultaneously decrease the formation of undesirable by-products (Al-Dhabi et al., 2020). For example, Al Dhabi *et al.* (Al-Dhabi et al., 2020), used municipal sludge combined with food waste. The co-fermentation increased LA concentration by 1.33-fold compared to mono-fermentation of municipal sludge.

## 4 Conclusions

Seaweed hydrolysate and a novel fermentation substrate consisting of candy-waste and digestate were compared as alternative feedstocks for LA production. Seaweed hydrolysate resulted in high relative LA production (85.21 ± 0.27%), but low lactic acid titer due to the initial sugar content of the seaweed. Up-scale experiments were performed with candy-waste: digestate (1:1) for 48 hours, producing 65.65 ± 2.05 g/L of LA, with productivity 1.37 g/L/hour. These findings promote the idea of studying the co-fermentation of industrial wastes, tackling simultaneously the issues of waste-management and sustainable product generation.

E-supplementary data of this work can be found in the online version of the paper.

**Acknowledgments**

This work was supported by the European Union's Horizon 2020 – Marie *Skłodowska-Curie* research and innovation program (grant agreement No 860477). The authors would like to thank the personnel of Analytical Services Unit of the Chemical Process and Energy Resources Institute, Centre for Research and Technology Hellas, Lidia Benedini and Hector Hernan Caro Garcia employees in Technical University of Denmark, for the analytical support. Furthermore, we would like to thank Ioannis V. Skiadas for the introduction to the bioreactor systems.